\documentclass[epsf]{aa}
\usepackage{astron}
\def\arcsec              {$^{\prime\prime}$}

\def\kmsmpc              {km\thinspace s$^{-1}$\thinspace Mpc$^{-1}$}

\begin{document}
%
%  These Macros are taken from the AAS TeX macro package version 4.0.
%  Include this file in your LaTeX source only if you are not using
%  the AAS TeX macro package and need to resolve the macro definitions
%  in the BibTeX entries returned by the ADS abstract service.
%
%  For more information on the AASTeX macro package, please see the URL
%	http://www.ferberts.com/AAS/aastex.html
%  For more information about ADS abstract server, please see the URL
%	http://adswww.harvard.edu/ads_abstracts.html
%

% Abbreviations for journals.  The object here is to provide authors
% with convenient shorthands for the most "popular" (often-cited)
% journals; the author can use these markup tags without being concerned
% about the exact form of the journal abbreviation, or its formatting.
% It is up to the keeper of the macros to make sure the macros expand
% to the proper text.  If macro package writers agree to all use the
% same TeX command name, authors only have to remember one thing, and
% the style file will take care of editorial preferences.  This also
% applies when a single journal decides to revamp its abbreviating
% scheme, as happened with the ApJ (Abt 1991).

% \let\jnl@style=\rm
% \def\rf@jnl#1{{\jnl@style#1}}
\def\rf@jnl#1{{#1}}
\def\aj{\rf@jnl{AJ}}                   % Astronomical Journal
\def\araa{\rf@jnl{ARA\&A}}             % Annual Review of Astron and Astrophys
\def\apj{\rf@jnl{ApJ}}                 % Astrophysical Journal
\def\apjl{\rf@jnl{ApJ}}                % Astrophysical Journal, Letters
\def\apjs{\rf@jnl{ApJS}}               % Astrophysical Journal, Supplement
\def\ao{\rf@jnl{Appl.~Opt.}}           % Applied Optics
\def\apss{\rf@jnl{Ap\&SS}}             % Astrophysics and Space Science
\def\aap{\rf@jnl{A\&A}}                % Astronomy and Astrophysics
\def\aapr{\rf@jnl{A\&A~Rev.}}          % Astronomy and Astrophysics Reviews
\def\aaps{\rf@jnl{A\&AS}}              % Astronomy and Astrophysics, Supplement
\def\azh{\rf@jnl{AZh}}                 % Astronomicheskii Zhurnal
\def\baas{\rf@jnl{BAAS}}               % Bulletin of the AAS
\def\jrasc{\rf@jnl{JRASC}}             % Journal of the RAS of Canada
\def\memras{\rf@jnl{MmRAS}}            % Memoirs of the RAS
\def\mnras{\rf@jnl{MNRAS}}             % Monthly Notices of the RAS
\def\pra{\rf@jnl{Phys.~Rev.~A}}        % Physical Review A: General Physics
\def\prb{\rf@jnl{Phys.~Rev.~B}}        % Physical Review B: Solid State
\def\prc{\rf@jnl{Phys.~Rev.~C}}        % Physical Review C
\def\prd{\rf@jnl{Phys.~Rev.~D}}        % Physical Review D
\def\pre{\rf@jnl{Phys.~Rev.~E}}        % Physical Review E
\def\prl{\rf@jnl{Phys.~Rev.~Lett.}}    % Physical Review Letters
\def\pasp{\rf@jnl{PASP}}               % Publications of the ASP
\def\pasj{\rf@jnl{PASJ}}               % Publications of the ASJ
\def\qjras{\rf@jnl{QJRAS}}             % Quarterly Journal of the RAS
\def\skytel{\rf@jnl{S\&T}}             % Sky and Telescope
\def\solphys{\rf@jnl{Sol.~Phys.}}      % Solar Physics
\def\sovast{\rf@jnl{Soviet~Ast.}}      % Soviet Astronomy
\def\ssr{\rf@jnl{Space~Sci.~Rev.}}     % Space Science Reviews
\def\zap{\rf@jnl{ZAp}}                 % Zeitschrift fuer Astrophysik
\def\nat{\rf@jnl{Nature}}              % Nature
\def\iaucirc{\rf@jnl{IAU~Circ.}}       % IAU Cirulars
\def\aplett{\rf@jnl{Astrophys.~Lett.}} % Astrophysics Letters
\def\apspr{\rf@jnl{Astrophys.~Space~Phys.~Res.}}
                % Astrophysics Space Physics Research
\def\bain{\rf@jnl{Bull.~Astron.~Inst.~Netherlands}} 
                % Bulletin Astronomical Institute of the Netherlands
\def\fcp{\rf@jnl{Fund.~Cosmic~Phys.}}  % Fundamental Cosmic Physics
\def\gca{\rf@jnl{Geochim.~Cosmochim.~Acta}}   % Geochimica Cosmochimica Acta
\def\grl{\rf@jnl{Geophys.~Res.~Lett.}} % Geophysics Research Letters
\def\jcp{\rf@jnl{J.~Chem.~Phys.}}      % Journal of Chemical Physics
\def\jgr{\rf@jnl{J.~Geophys.~Res.}}    % Journal of Geophysics Research
\def\jqsrt{\rf@jnl{J.~Quant.~Spec.~Radiat.~Transf.}}
                % Journal of Quantitiative Spectroscopy and Radiative Trasfer
\def\memsai{\rf@jnl{Mem.~Soc.~Astron.~Italiana}}
                % Mem. Societa Astronomica Italiana
\def\nphysa{\rf@jnl{Nucl.~Phys.~A}}   % Nuclear Physics A
\def\physrep{\rf@jnl{Phys.~Rep.}}   % Physics Reports
\def\physscr{\rf@jnl{Phys.~Scr}}   % Physica Scripta
\def\planss{\rf@jnl{Planet.~Space~Sci.}}   % Planetary Space Science
\def\procspie{\rf@jnl{Proc.~SPIE}}   % Proceedings of the SPIE

\let\astap=\aap
\let\apjlett=\apjl
\let\apjsupp=\apjs
\let\applopt=\ao

\input epsf   
\thesaurus{03(11.03.2; 11.04.2; 11.05.2; 11.06.1; 11.16.1; 11.19.3)}

\offprints{V.  Doublier
(ESO  address)} 
\title{POX~186: the ultracompact  Blue Compact Dwarf Galaxy
reveals its nature \thanks{based on observations carried out at NTT in La
Silla, operated by the  European Southern Observatory, during Director's
Discretionary Time.}}
\author{V. Doublier\inst{1,2}
\and D. Kunth\inst{3}
\and F. Courbin\inst{4,5}
\and P. Magain\inst{4}
}
\institute{European Southern Observatory, Alonso de Cordova 3107,
Casilla 19001, Santiago, Chile
\and Observatoire de Marseille and Institut Cassendi, 2 place Le
Verrier, F-13004 Marseille, France
\and Institut d'Astrophysique de Paris, 98bis Bld Arago, F-75014 Paris,
France
\and Institut d'Astrophysique, Universit\'e de Li\`ege,
Avenue de Cointe 5, B--4000 Li\`ege, Belgium
\and URA 173 CNRS-DAEC, Observatoire de Paris, F--92195 Meudon
Principal C\'edex, France}
\date{Accepted 11/02/99}

\titlerunning{POX~186}
\authorrunning{Doublier et al.}
\maketitle

\begin{abstract}

High  resolution, ground based $R$   and $I$ band observations of  the
ultra compact dwarf galaxy  POX~186 are presented.  The data, obtained
with the ESO New Technology Telescope (NTT),  are analyzed using a new
deconvolution algorithm which allows  one   to resolve the   innermost
regions of  this stellar-like  object into  three  Super-Star Clusters
(SSC).   Upper limits to  both masses (M$\sim$ 10$^5$ M$_{\odot}$) and
the physical sizes ($\le$60pc) of  the SSCs are  set. In addition, and
maybe most importantly, extended light emission underlying the compact
star-forming region is clearly detected in both bands. The $R-I$ color
rules out nebular H$\alpha$ contamination and is consistent with an old
stellar  population.   This casts  doubt on the   hypothesis that Blue
Compact Dwarf Galaxies (BCDG) are young galaxies.


\end{abstract}

\section{Introduction}

Among the Blue Compact Dwarf  Galaxy (BCDG) population, there exists a
small class  of  very  compact  star-forming objects,   unresolved  on
photographic and {\rm CCDs} images.   This small fraction ($<10$\%) of
BCDGs   does  not appear to   possess  {\em   any} underlying  stellar
component substantially  older    than the    starburst     population
(\cite*{thuan83};   \cite*{kunthetal88};   \cite*{drinkwaterhardy91}).
The most representative galaxy of this  kind is the star-forming dwarf
POX~186  (RA:13h23m12.0s;  DEC:-11$^{\circ}$       22'; B1950).   This
relatively nearby galaxy (recession velocity $v=1170$ km s$^{-1}$) was
discovered in  a  prism-objective survey search (\cite*{kunthetal81}),
and  has been  observed spectroscopically as   part of a primordial He
abundance study (\cite*{kunthsargent83}). Ground-based images taken at
ESO  (\cite*{kunthetal88})  revealed  a nearly round,  barely resolved
object with no apparent substructure.   The luminosity of this  galaxy
is about  $M_V =  -14$ (\cite*{kunthetal88}) and  population synthesis
models predict that it should have a  post-burst luminosity as low as $M_V
= -10$  .  Therefore, this object  may represent the  smallest unit of
galaxy formation.  Given POX~186's luminosity, its total luminous mass
should   be 10$^8$M$_{\odot}$ ($H_{0}$=80   \kmsmpc\ and mass-to-light
ratio of 1.). This exceeds by 1  to 2 order  of magnitudes the mass of
star clusters observed in our Galaxy or in neighboring galaxies.

It has been conjectured that POX~186 could be a young, isolated, giant
star-forming  cluster.  The advent of   HST, with its very high
angular   resolution,  has   permitted  the    discovery of  so-called
``Super-Star  Clusters'' (SSC)  found  in many  star-forming dwarf and
irregular galaxies.  One of the  first and most remarkable examples is
He2-10 by \cite{contivacca94}, where UV images obtained with the Faint
Object Camera on HST revealed 10 bright  star clusters near the center
of  the galaxy as well   as a large number  of  fainter clusters in  a
starburst region $8''$ (350 pc) away from the center.

In this    paper, we present  optical  observations  obtained with the
New Technology Telescope (NTT) at ESO, La Silla (Chile), and report on
the  presence  of an extended   faint  halo  underlying the  starburst
component.  We  show also that,  using   deconvolution, POX~186 can be
resolved into several substructures which can be explained in terms of
SSCs.

\begin{figure*}[t!]
\begin{center}
\leavevmode
\epsfxsize=12cm
%\picplace{12.0cm}
\epsffile{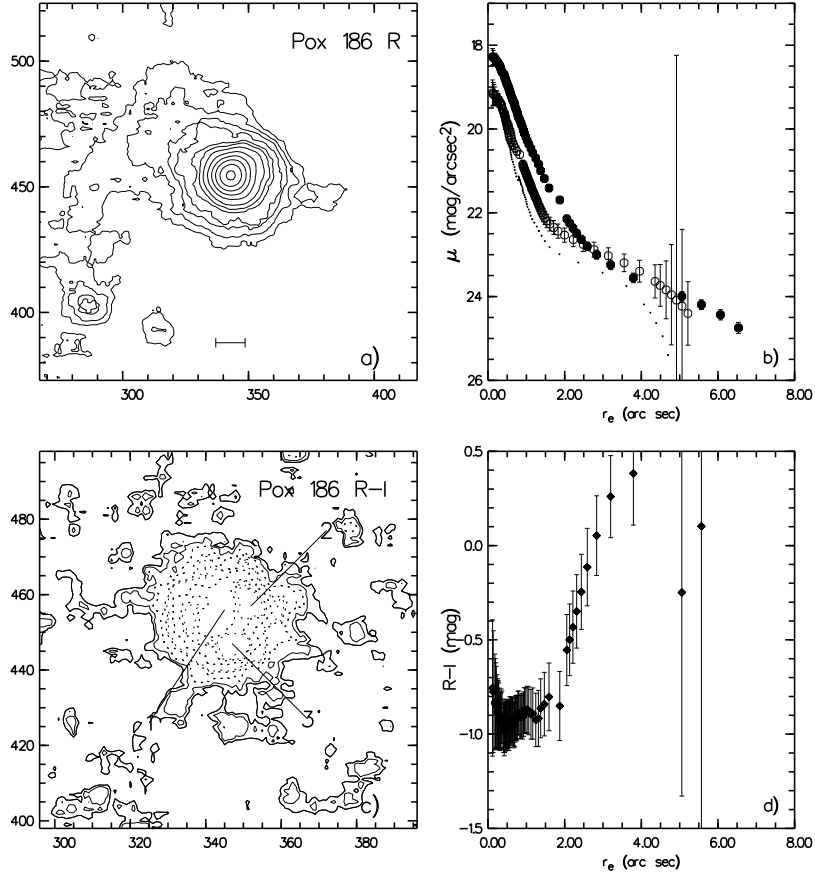}
\end{center}
\caption{Surface photometry of POX~186.}  {\bf a)} An $R$ band contour
plot of  POX~186, the size of the field is 
18\arcsec$\times$18\arcsec, North is up, and East is right. The contour
interval is 0.5 mag.. The lowest
contour corresponds to a surface brightness of 24 mag arcsec$^{-2}$.
The bar corresponds to a
linear size of  100 pc (the linear  scale  is 71 pc/arcsec).  {\bf
b)} The  light distribution profiles in  $R$ (filled circles)
and $I$ (open circles).  The error  bars on the light
profiles  have been  computed assuming that Poisson  statistics apply.  The
dotted  profile  corresponds to  the  $R$  profile of a nearby non-saturated
star.   {\bf 
c)} A Contour plot of the $R-I$ map. The size of the field is 12$''
\times$12$''$. The  {\bf lowest} contour corresponds to a $R-I$ 
color of --0.8 mag  {\bf in the center  of the galaxy}; the intervals between
the contours  are 0.1 mag, and  the dashed contours represent negative
values. {\bf d)} The  $R-I$ color profiles,  uncorrected  for Galactic
extinction. The errors are computed from the errors on the $R$ and $I$
profiles.  Note the presence of 3 ``blobs'' on the $R-I$ map.  

\end{figure*}

\section{NTT observations}

POX~186 was observed in the $R$ and $I$ bands at the NTT (ESO, Chile)
with the SUperb Seeing Imager (SuSI) in February 1996.  The pixel size
of the detector is 0.128$''$.  The
weather conditions were photometric  and the seeing was better than
0.9$''$. The exposures were split into  two 10-minute exposures due to
the high sky background.

The  first part  of the data  reduction (i.e., bias subtraction and
flat fielding using dome and  sky flat-fields) was  performed using the IRAF
package. The cosmic-rays removal procedure  available in MIDAS was
applied to the images.  As the  scattered light from  the moon caused
an uneven  sky-background  around the
galaxy, the    sky was subtracted with a frame obtained  by  smoothing
the  reduced   images with a
median filter and a large smoothing box,  while masking the objects in
the frames.  This  was applied to  each of the  two frames  which were
then averaged. 

\subsection{Surface photometry}

The photometric results are summarized in Table 1. This includes the
asymptotic magnitude  m$_{\mathrm{asym}}$  (extrapolated integrated magnitude),
the effective   surface brightness $\mu_{\mathrm{eff}}$  in
mag/arcsec$^2$ (surface 
brightness  at the half light  radius), the effective radius
r$_{\mathrm{eff}}$ in 
arcseconds (half light  radius) and the effective  mean surface brightness
$<\mu_{\mathrm{eff}}>$    (mag/arcsec$^2$)   (mean  surface   brightness within
r$_{\mathrm{eff}}$),  for  the $R$  and  $I$   bands.   Finally, the
scalelength parameter ($\alpha$) and the central surface brightness
($\mu_0$) derived by extrapolating the fit at the equivalent radius to
zero, are given.

The errors on the integrated magnitudes are 0.05 mag in both the $R$
and $I$   bands.  Errors on  the  surface brightness  measurement were
computed assuming that Poisson statistics applied.

\begin{table}[t]
\begin{center}
\caption{Photometric results for POX 186}
\begin{tabular}{l | c c | c c}
\hline \hline
& R & & I & \\
\hline
m$_{asym}$ & 16.36 & & 17.05 &\\
$\mu_{eff}$ & 22.23 & & 21.52 & \\
r$_{eff}$ & 3.75 & & 2.35 & \\
$<\mu_{eff}>$ & 21.20 & & 20.90 & \\
$\mu_0$ (exp) & 22.03 (0.04) & & 21.68 (0.01*) & \\
$\mu_0$ (r$^{1/4}$) & 13.52 (0.21) & & 13.23 (0.08) & \\
$\alpha$ (kpc$^{-1}$) & 5.56 (0.72) & & 6.14 (0.40)& \\  
\hline
\end{tabular}
\end{center}
(*): 1-$\sigma$ error\\

\end{table}

A 18\arcsec  $\times$18\arcsec  $R$ band contour plot
centered on POX~186 is shown in Fig. 1a. The bar indicates a linear
size of 100 $pc$ (H$_0$ =
80 \kmsmpc).   The surface brightness     profiles in the  $R$ band
(filled
circles) and $I$ band (open  circles) are displayed in Fig. 1b.  The
$R-I$ color  map   (12$'' \times$12$''$), and the color 
distribution profile  as   a    function  of  the equivalent
radius
(\cite*{devaucouleursaguero73}) are displayed in Figs. 1c and 1d, respectively.

A more detailed  explanation of  the   surface  photometry  and  color
distribution profiles is given in \cite{doublieretal97}.

\subsection{Images and Light distribution profiles}

POX~186 is clearly resolved (see  Fig. 1a) in our sub-arcsecond images
(r$_{eff}$(R)  =   3.75$''$,  r$_{eff}$(I)   = 2.35$''$).  An  obvious
asymmetry in  the external isophotes can  be seen on
the Eastern side of the nucleus. It is seen in both our $R$ and $I$
band images.  Two
different  regions  are  seen on   the  $R-I$ color  plot (fig. 1d).
While the
innermost region   of the galaxy is blue   ($R-I\sim -0.8$), the outer
part shows a red color gradient. Over a region extending from 2$''$ to
6$''$  (i.e., over about 280 pc)  away from  the nucleus,  the $R-I$ color
increases from  --0.8 mag to 0.4  mag.  This  color distribution
translates into a color gradient of $\sim4.2$ mag/kpc.

Furthermore, the $R-I$ map (Fig. 1b) reveals three distinct
substructures in the
central parts of the  galaxy (labelled ``1'', ``2'', ``3''), the
largest of  them being centered with
respect  to  the   outer isophotes.   This  suggests  that the central
nucleus, which  was originally thought to be single, consists of smaller
units.

Alternatively,   the   observed structures  could      be due to signal
fluctuations  due to  poor  alignment of the frames.  Such an effect
would be enhanced when creating the color maps.   However, other
objects (stars) do  not show  such
artifacts.  We can therefore safely consider these structures to be
real and intrinsic to the galaxy.

We used the algorithm   developed by \cite{sagliaetal97}, to  fit the
light distribution profiles in  the  $R$ and  $I$ bands. The   central
parts of the two profiles are not well  fitted by a r$^{1/4}$ law. A
model that includes an exponential profile and an r$^{1/4}$ gives a
better fit to the data; however, the best fit still leaves significant
residuals
in the center of  the frame. This suggests  that POX~186 is not  a single
spheroidal object,  like a large  star cluster.  On  the contrary, the
galaxy appears to be composed of a nucleus with a profile broader than
a  r$^{1/4}$  law,  plus    underlying  emission  compatible   with an
exponential profile.

\section{Underlying galaxy and starburst}

The  extended   features observed in   the  $R$  and  $I$  frames  and
``knotty''   center of   the     $R-I$ map  have been deconvolved with
a new deconvolution  algorithm developed by
\cite{magainetal98}.  The algorithm decomposes the  images into a sum
of point sources plus a diffuse deconvolved background smoothed on a
length scale  {\it chosen by  the user}.  As 
output, the program  returns the photometry  and position of the point
sources, as well as a deconvolved image.

The  main idea  behind  the new  deconvolution technique, is  to avoid
deconvolution by the total Point  Spread Function (PSF) of the  image,
Instead, a narrower function is used,  allowing  the
spatial frequencies of the deconvolved image to stay within the limits
imposed  by  the sampling  theorem.   As a   consequence, most of  the
so-called ``deconvolution  artifacts''   are  avoided.  

The good sampling of  the  images  (pixel size of 0.128\arcsec) and
the good stability  of the PSF across  the  field make our data  well
suited for deconvolution.  The FWHM in the undeconvolved images is of the
order of  0.85\arcsec. The deconvolved images have a resolution of
0.38$''$. This is dictated by the signal-to-noise of our data.  The
PSF needed for deconvolution was obtained
from a star, about as bright  as POX~186 and situated 1$'$ away.

\begin{figure}[t]
%\begin{center}
\leavevmode
\epsfxsize=7.5cm
\epsffile{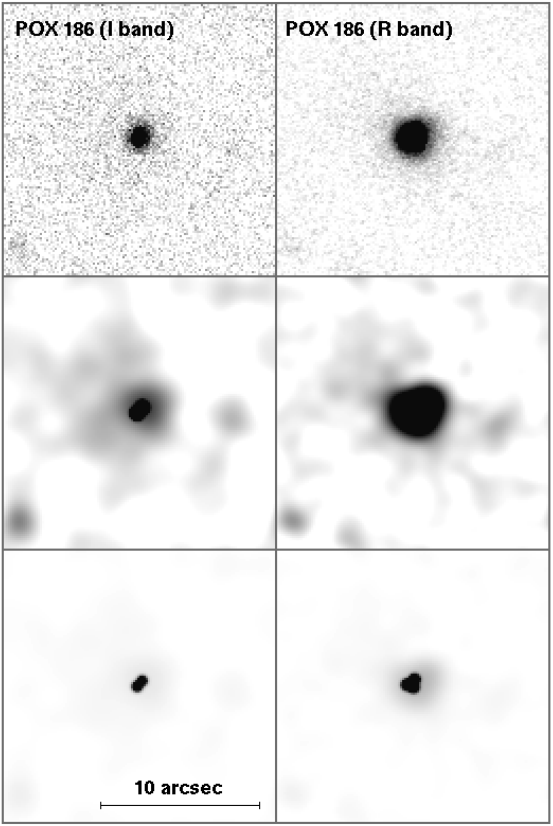}
\caption{Top panels show the reduced  $R$ and $I$  band frames. Middle 
and bottom  panels show  the  result of   the  deconvolution with
different intensity cuts. }
%\end{center}
\end{figure}

We consider that the solution  found by the deconvolution procedure is
good when the residual map,  in units of  the photon noise, has the
correct statistical distribution across the whole field, i.e. equal to
one (or  very close, see \cite{courbinetal98} and
\cite{burudetal98} for more details on this procedure).

The residual  maps obtained show significant  structures when only one
point source is used to  model the center  of POX~186.  A second point
source was added to obtain good residual maps in the $I$ band, whereas
a third source had to be used to fit properly POX~186's nucleus in the
$R$ band.  Note  that adding more point  sources leads to over-fitting
the central parts  of the galaxy.  We  can therefore conclude that the
present data are compatible with a galaxy nucleus  composed of 3 point
sources in $R$ and at least 2 in $I$.

Figure 2 shows  the $R$ and $I$  band images with different cut-off intensity
in order to display better either the faint  extensions of the galaxy,
or the bright central region.  While the diffuse halo of the galaxy was
already detected in  the reduced $R$  band frame (S/N   $>$ 3), it  is
hardly   seen in the   $I$  frame (S/N $\le$   3)  despite  an obvious
asymmetry of the  image.  This diffuse  halo clearly  shows up on  the
surface brightness profiles, where  we can reach brightness levels  as
faint as a few percent of the sky background level.  The diffuse halo
can also be seen in the deconvolved images.

\begin{table}[t]
\caption{Derived luminosities and masses of the deconvolved clusters}
\begin{center}
\begin{tabular}{c | c c c| c c c}
\hline \hline
&&R&&&I& \\
N$^{o}$ & m & M$_R$ & Mass* & m & M$_I$ & Mass* \\
\hline
1 & 19.23 & -11.59 & 2.3 & 19.61 & -11.21 &1.2 \\
2 & 19.27 & -11.55 & 2.2 & 19.85 & -10.97 & 0.9 \\
3 & 19.47 & -11.35 & 1.8 & not &detected &- \\
\hline
\end{tabular}
\end{center}
(*): Mass in units of 10$^6$ M$_\odot$ \\
\end{table}

From the deconvolved images, we obtained an estimate of the luminosity
and the masses of the ``clusters'' (Table 2).   Because of the limited
S/N  ratio and the heavy blending  of  the knots, the magnitude errors
are  large, about 0.5  mag.   The absolute  magnitudes are  computed
assuming  $H_{0}$  =  80 \kmsmpc,  and  the  masses  are  determined by
assuming a mass-to-light ratio of 1 (recent HI observations of POX~186
set  an  upper limit  of a  few  10$^6$ M$_\odot$,  S. Cot\'e, private
communication).  The  masses are   large  compared with  the  observed
masses   of  single SSCs   (\cite*{meureretal95}).   However,   if the
mass-to-light ratio  estimates of  \cite{charlotetal96}  for  a single
burst stellar  population of age  less than  10$^7$ years ($M/L  = 0.1
M_{\odot}/L_{\odot}$) are more realistic,  the ``clusters'' in POX~186
will have  smaller masses   (10$^4$ to  10$^5$  M$_{\odot}$). This  is
similar to those of SSCs observed in other starburst galaxies.

The physical sizes of the knots are difficult to ascertain. They  are
unresolved and the only  constraint we can set  on their size is given
by the seeing of the  observations, 0.85$''$, or  about 60 pc.   Their
size on  the deconvolved image,  0.384$''$ (about 26 pc) only reflects
the adopted resolution limit.

\begin{figure}
\begin{center}
\leavevmode
\epsfxsize=8.5cm
\epsffile{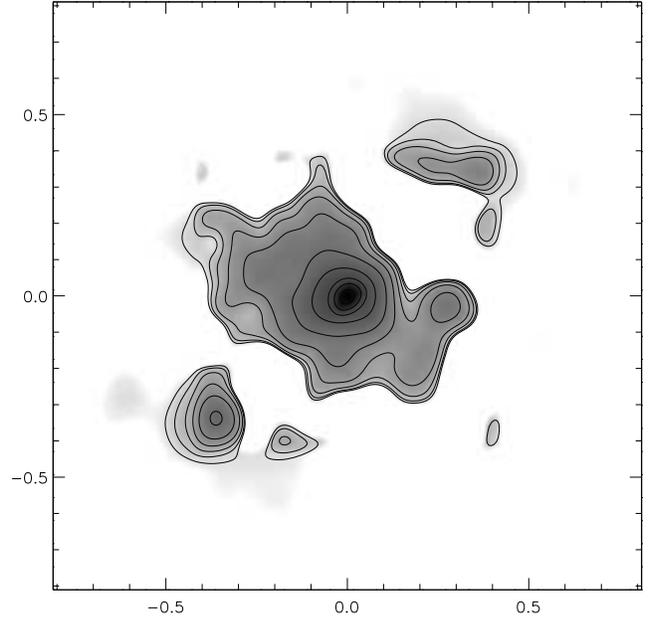}
\caption{Two-dimensional cross-correlation  map  for the  $R$ and  $I$
frames. The grey levels  increase from 0.  (no correlation,  white) to
1.  (exact correspondence, black). The contours are levels above
0.5, and the step between contours is 0.05}
\end{center}
\end{figure}

As a check on the accuracy of the deconvolution, we  also generated the
two-dimensional   cross-correlation  map  of  the  $R$   and  $I$ band
``un-deconvolved''  frames (Fig.  3).   Only  the levels above 0.5 are
displayed in Fig.  3 (a level of 0. implies no correlation and a level
of 1. implies full correlation).   The cross-correlated
image  reproduces very well the  structures  seen in both  deconvolved
frames. This indicates that  they are real and that they are not due to 
noise enhancement  introduced by  poor   sky subtraction, or  by   the
deconvolution process.

One of the knots in $R$ is  not detected in the $I$ band,
although the cross-correlation map shows an obvious asymmetry at the
corresponding position on  the $R$ frame. The residual
maps after  deconvolution in $R$,  are significantly better when three
point source  are considered.  Detecting only  two of  them in the $I$
band may therefore indicate that the third  is too faint.  Indeed, the
$I$ band magnitude of the counterpart  of the ``missing'' knot  would
be about 20 mag. which
is still consistent with the central $R-I$ index of --0.8.

The   extended component  of POX~186  shows   a filamentary  structure
instead of   a smooth  light distribution  as  usually seen  in  BCDGs
(\cite*{papaderosetal96},                        \cite*{tellesetal97},
\cite*{doublieretal97}).   Although this is at  the limit of the noise
level  of the  present  data,  the  filaments  can   be seen in both
the deconvolved images and the cross-correlation map. The low surface
brightness component of POX~186 may  be similar  to
the filamentary  structures seen in dwarf  irregulars, rather  than
the smoother surface profiles of BCDGs.

Further observations   in  the near-IR  with high-sensitivity
detectors  and larger  telescopes (the  expected 
surface brightness of an  evolved  stellar population  in $K$ band  is
below 21   mag. arcsec$^{-2}$,  i.e.,  the   limit  for  4  meter-class
telescope with  the  current  detectors)  will elucidate the
nature of the stellar population in this ``halo''.

\section{Discussion}

\subsection{Nature of the galaxy}

%We have shown that the extended component seen around the nucleus consists of
%a real  underlying faint   extended  structure,  rather than  of    an
%unrealistic extra component necessary to represent the data.

We have shown that the faint halo seen around the nucleus of POX~186
is a real structure, rather than an artefact produced by sky
background subtraction, or by the deconvolution.

The  nature of this diffuse component could be  due to an
evolved stellar  population, or to H$\alpha$ emission contaminating
the $R$ band, leaking  from the star
forming region. However, if   the  extension was  due  to  H$\alpha$
emission,    the  $I$ band  profile would not be so similar. Moreover,
the $R-I$ color distribution would not
exhibit  a steep color gradient. Its  observed apparent increase is in
fact due to a  decrease of the contribution of  the $R$ band flux with
respect  to the $I$  band. This is definitely  not consistent with the
contamination hypothesis of the $R$ band by H$\alpha$ at large radii.

%On one hand, the  asymptotic $R-I$ color,  not corrected for  Galactic
%extinction or   internal reddening, is $-$0.70   (while  $B-R$ = 1.28,
%\cite*{kunthetal88}).   It  is   completely dominated by   the central
%starburst  and the diffuse  emission does  not  contribute much to the
%``blue colors'',  unlike  in  most BCDGs   (where  the underlying  old
%population    dominates at  large    radii).  The  equivalent width of
%H$\beta$ (370 \AA; \cite*{kunthsargent83}) suggests that the starburst
%cannot  be    older  than    5    Myrs    (\cite*{leithererheckman95};
%\cite*{mashessekunth91}).   Thus the nebular emission lines contribute
%largely to the colors in the central regions.
 From the
previous observations, no underlying component was detected, and it
was argued that the red colors obtained by \cite*{kunthetal88} 
($B-R$ = 1.28) were
due to H$\alpha$ emission contaminating the $R$ band filter. However, 
together with
the $B-R$ color, the $R-I$ color of the faint extension  is consistent  with
emission  dominated by    red  giant  stars    of spectral type G
(\cite*{johnson66}).  From the observed  emission-line ratios, we know
that internal  reddening  is negligible,  so that  the $R-I$ color
cannot be attributed to dust.   There is therefore strong evidence for
an extended faint component of old stars underlying a starburst region.

\subsection{POX~186 among the other dwarf galaxies population}

POX~186 is a very small and compact  object compared with other BCDGs:
its effective radius is less  than 300 $pc$ in $R$,  and less than 200
$pc$ in $I$, while the mean effective  radius ($R$ band) for a typical
BCDG      is     about      800    $pc$      (\cite*{doublieretal97};
\cite*{doublieretal99}).    The  compactness   index (ratio of the
effective radius to the radius at 1/4th of the total luminosity) of
POX~186  is
r$_{0.5}$/r$_{0.25}  = 3.3$ in  $R$;  the mean compactness index value
for BCDGs is 2.3  (\cite*{doublieretal99}). It is  1.78 in the $I$
band.  This indicates that the galaxy is much less compact in $I$ than
in $R$,  supporting  the hypothesis  of the  presence of   an extended
structure underlying the starburst component.  This compares well with
irregular galaxies which have  compactness   indexes of 1.6,    whereas
elliptical galaxies    have  larger  compactness  indexes   of 2.9-3.0
(\cite*{fraser77}; \cite*{devaucouleursaguero73}).

Using the $B-R$ color given by \cite*{kunthetal88} to estimate a
lower limit  for the central surface  brightness $\mu_0$(B) in the $B$
band, we  can place  POX~186 in  the diagram [$\mu_0$,  M$_B$] derived by
\cite{binggelicameron91}.  POX~186 is located in the
continuation of  the sequence defined by  elliptical galaxies and
spiral bulges.  The  exponential underlying component is   located
within  the  sequence defined by    the dwarf irregulars  and disks  of
spirals in the    [$\mu_0$, M$_B$] plane.   In  the [$\log{(\alpha)}$,
M$_B$] diagram, the underlying component falls well  below the bulk of
dEs, dIS and S0s.  It is therefore more likely to be associated  to  Low
Surface  Brightness  galaxies than to  dwarf irregulars.

\begin{figure*}[!t]
\begin{center}
\leavevmode
\epsfxsize=12cm
\epsffile{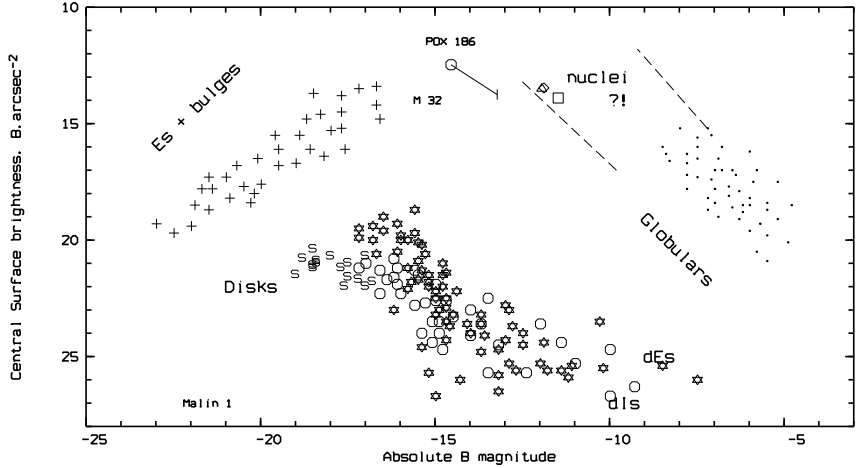}
\caption{[$\mu_0$, M$_B$] plane: elliptical galaxies and bulges of
spirals are represented as crosses, spiral disks as ``s'', dwarf
galaxies as open symbols: stars for dwarf ellipticals and open circles
for dwarf irregulars, globular clusters are represented as points. POX
186 is placed twice: the open circle corresponds to [$\mu_0$, M$_B$]
for total galaxy, the triangle and the square represent the clusters
detected  in the deconvolved images.}
\end{center}
\end{figure*}

The  3 ``clusters'' lie well  within  the ``galactic nuclei''  sequence
discussed by  \cite{binggelicameron91} and \cite{phillipsetal96}, in
the [$\mu_0$, M$_B$] plane. In addition,  population synthesis
evolutionary models predict that the star forming regions fade
considerably following an instantaneous burst (about 3-5 magnitude in the B
band; \cite*{cervinomashesse94}). If we apply the predicted dimming
to the 3 clusters, they evolve to the Globular Cluster
sequence in the [$\mu_0$, M$_B$] plane.

%Moreover, dimming the luminosity of the 3
%``clusters'' using population synthesis evolutionary models would
%allow them  to fall right into the ``globular clusters'' sequence.

\section{Conclusion}

The primary  result of  our  study is  the  detection of  an
extended  faint component  of  old stars .

As revealed  by optical CCD imaging, most  BCDGs show a faint envelope
surrounding the compact central
starburst   region    (\cite*{thuan83};   \cite*{hunter1gallagher85};
\cite*{kunthetal88}).   Near-infrared   observations and  evolutionary
synthesis  models have shown that these  faint envelopes are composed
of   stars      older     than  several     Gyrs     (\cite*{thuan83};
\cite*{hunter1gallagher85}; \cite*{doublieretal99}).  However, any previous
major episode of star formation must  have been  rather  mild  since  the
large value of the
H$\beta$ equivalent   width (370\AA) observed   in the very  center of
POX~186  (\cite*{kunthsargent83})   indicates  that   the   underlying
component does not contribute enough to the optical continuum so as to
dilute   significantly  the  emission  line.    Moreover,  the optical
spectrum shows a  very   high  excitation and the    overall  measured
metallicity is low  ($Z \sim Z_\odot/10$, \cite*{kunthsargent83}).
Using a closed-box  model for chemical  enrichment, the  present burst
could account for most of  the metal enrichment
and continuum light  (\cite*{marconietal94}).  No HI has been detected
in POX~186. It could be that it has been exhausted. Alternatively, the
gas could have
been highly ionized and/or dispersed by the strong  winds generated by the
massive stars in the clusters.

Our second result is the discovery of Super Star Clusters (SSCs) in
POX~186.  The
deconvolution technique   reveals the    presence  of at    least  two
``clusters'' in the center of POX~186, with properties similar to that
of  the  SSCs  observed  by  \cite{meureretal95}  in other  starburst
galaxies.  They  found SSCs in HST FOC  UV images of a
sample of nine  starburst galaxies. The SSCs  in their sample  account
for about 20\% of the total observed  UV flux.  Similar SSCs have been
discovered  in several other irregular or starburst
galaxies (e.g., NGC  1569, NGC 1705: \cite*{oconnelletal94}; NGC 1140:
\cite*{hunteretal94};    M82: \cite*{oconnelletal95};  NGC       4214:
\cite*{leithereretal96};  NGC     253:  \cite*{watsonetal96}).   These
clusters have luminosities  in the  range $-14<M_V<-11$, diameters  of
the order of  10 pc or less,  and  estimated masses in the  range
$10^4  -     10^6~M_{\odot}$      (e.g.,         \cite*{hunteretal94};
\cite*{oconnelletal94};   \cite*{contivacca94}).  Typical   sizes  and
upper-mass estimates  of SSCs are  consistent  with those of  Galactic
globular clusters and it is suggested  that these SSCs may be globular
clusters in the  process of formation, with  ages of less than  $10^7$
yrs (\cite*{contivacca94}).  At this point, we remark upon the work by
\cite{phillipsetal96} based on the nuclei  and ``knots'' of late-type
galaxies. They find an obvious overlap in properties between the large
star  clusters in these galaxies and  the nuclei.  Current discussions
aim at establishing whether stellar-like  cusps could be remnants of a
previous starburst  episode  (\cite*{kormendy89a}).  We believe that
the multiple sources observed in POX~186 could be  SSCs like in other
starbursting galaxies.   Their derived  absolute magnitudes agree well
with those of  SSCs  observed in starburst  galaxies.   Clearly, it is
mandatory to confirm the   existence of the substructures  in POX~186:
SSCs and underlying    galaxy.   First, POX~186 remains one    of  the
smallest star-forming galaxy: the  exponential scale length of POX~186
( $<$ 200 pc  ) is small compared to other known BCGDs. Second, the
presence of  SSCs raises  new questions  about its   future evolution.
Finally, the presence of the underlying  old component removes POX~186
from the ``young'' galaxy candidates list.

\begin{acknowledgements}
Part of this work was performed during Daniel  Kunth's visit at ESO in
a   Visiting    Astronomer  program.  The   authors  are   thankful to
M. Mas-Hesse for the  very helpful discussion.  VD wishes to thank the
Institut  d'Astrophysique de Paris  where  a good deal  of the present
work was done.  FC is supported by ARC 94/99-178 ``Action de Recherche
Concert\'ee de la Communaut\'e Fran\c{c}aise'' and P\^ole d'Attraction
Interuniversitaire  P4/05  (SSTC,  Belgium).  We   wish  to thank  the
anonymous  referee.
\end{acknowledgements}

\bibliographystyle{astron}

\bibliography{ref_p}

\end{document}